\documentclass[12pt,a4paper]{article}

\usepackage{epsf}

\newcommand{\eqpt}{\hspace{6pt}.}           
\newcommand{\eqcm}{\hspace{6pt},}

\newcommand{\eqref}[1]{(\ref{#1})}          

\newcommand{\gsim}{\raisebox{-.5ex}{\footnotesize$
     \,\:\stackrel{\textstyle>}{\sim}\,\:$}}    

\newcommand{\gp}{\gamma^{\ast} p}           
\newcommand{\pom}{{I \! \! P}}              

\newcommand{\Mx}{M}                         
\newcommand{\mq}{m_c}
\newcommand{\Pt}{P_T}
\newcommand{\lt}{l}

\newcommand{\iv}{v}
\newcommand{\iw}{w}

\newcommand{\GeV}{\mbox{\ GeV}}             
\newcommand{\pbarn}{\mbox{\ pb}}
\newcommand{\nbarn}{\mbox{\ nb}}

\begin{document}

\pagestyle{empty}
\setcounter{page}{0}
\begin{flushright}
{CPTH-S492-0197}
\end{flushright}
\vspace{\baselineskip}
\begin{center}
{\LARGE OPEN CHARM PRODUCTION\\
\bigskip
IN DIFFRACTIVE $ep$ COLLISIONS\\ }
\vspace{3\baselineskip}
M. Diehl\footnote{email: diehl@orphee.polytechnique.fr} \\
\vspace{\baselineskip}
\textit{Centre de Physique Th\'eorique\footnote{Unit\'e propre 14 du CNRS} \\
  Ecole Polytechnique, 91128 Palaiseau Cedex, France}
\vspace{6\baselineskip}
\begin{abstract}
  The cross section for the diffractive reaction $\gamma^\ast + p \to
  c \bar{c} + p$ with a real or virtual photon is calculated in the
  nonperturbative two-gluon exchange model of Landshoff and Nachtmann.
  Numerical predictions are given for cross sections and spectra at
  typical HERA values of c.m.\ energy and photon virtuality. The
  contribution of charm to the diffractive structure function is
  evaluated and found to be rather small in the model, and the ratio
  between the production rates for $b \bar{b}$ and $c \bar{c}$ is
  tiny.
\end{abstract}
\end{center}
\newpage
\pagestyle{plain}

\section{Introduction}
\label{sec:intro}

The discovery of diffractive events in $ep$ collisions at HERA
\cite{HERA:discover} has triggered a large amount of experimental and
theoretical work and greatly increased our knowledge of the physics of
diffraction, or of the pomeron. Several models give a reasonable
description of the data at the time being, but we are far yet from a
coherent picture of the mechanisms at work in terms of QCD. One can
hope that the detailed study of the diffractive final state will lead
to further progress in this direction. Charm production looks
promising in this respect, as predictions for this process differ
widely between various models
\cite{HERA:workshop,GehrStir,Nikolaev,Teubner,Lotter}.

In this paper we use the approach due to Landshoff and Nachtmann (LN)
to model the soft pomeron by the exchange of two nonperturbative
gluons. We present differential cross sections for the diffractive
dissociation of a real or virtual photon into a $c \bar{c}$-pair.
Provided that the invariant mass of the diffractive final state is not
too large its $c \bar{c}$-component should give a fair approximation
of inclusive diffractive charm production. In the following section we
give some details of the model and of the calculation, in
sec.~\ref{sec:results} we present our results, and in
sec.~\ref{sec:sum} we summarise our findings.

\section{Diffractive $c \bar{c}$-production in the LN model}
\label{sec:calc}

The Landshoff-Nachtmann model has been introduced and described in
\cite{LN,DL}. It approximates the soft pomeron by the exchange of two
\emph{nonperturbative} gluons, with a propagator $- g_{\mu \nu}
D(l^2)$ instead of the perturbative $- g_{\mu \nu} / l^2$ in Feynman
gauge. In several processes one can express the scattering amplitude
in terms of a few moments of the function $D(l^2)$ and thus does not
need to know its detailed form. Here we only need the integral
\begin{equation}
  \label{moment}
  \int_{0}^{\infty} d l^{2} [\alpha_{s}^{(0)} D(-l^{2})]^{2} 
  \cdot l^{2} = \frac{9 \beta_{0}^{2} \mu_{0}^{2}}{8\pi}  \eqcm
\end{equation}
where $\beta_{0} \approx 2.0 \, {\rm GeV}^{-1}$ and $\mu_{0} \approx
1.1 \, {\rm GeV}$ have been extracted from experimental data
\cite{DL,Cudell}. The parameter $\mu_0^2$ provides the characteristic
scale for the dependence of $D(l^2)$ on the gluon virtuality $l^2$,
and $\alpha_{s}^{(0)}$ stands for the strong coupling in the
nonperturbative region which dominates the $l^2$-integration in
\eqref{moment}. It will be taken as $\alpha_{s}^{(0)} \approx 1$ here
\cite{Cudell}.

We now apply this model to the reaction
\begin{equation}
  \label{reaction}
  \gamma^\ast + p \to c \bar{c} + p  \eqpt
\end{equation}
In the following we use the conventional variables $x, y, s, t, Q^2,
W^2$ for deep inelastic scattering, $\Mx$ for the invariant mass of
the $c \bar{c}$-pair, and
\begin{equation}
  \label{standard}
  \beta = \frac{Q^2}{Q^2 + \Mx^2 - t} \eqcm \hspace{4em}
  \xi = \frac{Q^2 + \Mx^2 - t}{W^2 + Q^2}  \eqpt
\end{equation}
We denote with $\Pt$ the transverse momentum of the charm quark with
respect to the photon momentum in the $\gp$ c.m.

\begin{figure}
  \begin{center}
    \leavevmode
    
    \setlength{\unitlength}{6truemm}
    \begin{picture}(20,10)(1,-1)
      \put(0,0){
        \begin{picture}(9,10)
          \thicklines
          \put(8,6.5){\oval(10,2)[l]}
          \put(0,0){\line(1,0){8}}
          \multiput(4,0)(0,1){6}{\line(0,1){0.5}}
          \multiput(6.5,0)(0,1){6}{\line(0,1){0.5}}
          \multiput(.25,6.5)(1,0){3}{\oval(0.5,0.5)[t]}
          \multiput(0.75,6.5)(1,0){3}{\oval(0.5,0.5)[b]}
          \put(4.75,7.5){\vector(1,0){1}}
          \put(0.75,0){\vector(1,0){1}}
          \put(1.5,7.3){$\gamma^\ast$}
          \put(8.3,7.3){$c$}
          \put(8.3,5.2){$\bar{c}$}
          \put(3.5,-1.5){$(a)$}
        \end{picture}
        }
      \put(12,0){
        \begin{picture}(9,10)
          \thicklines
          \put(8,6.5){\oval(10,2)[l]}
          \put(0,0){\line(1,0){8}}
          \multiput(4,0)(0,1){8}{\line(0,1){0.5}}
          \multiput(6.5,0)(0,1){6}{\line(0,1){0.5}}
          \multiput(.25,6.5)(1,0){3}{\oval(0.5,0.5)[t]}
          \multiput(0.75,6.5)(1,0){3}{\oval(0.5,0.5)[b]}
          \put(4.75,7.5){\vector(1,0){1}}
          \put(0.75,0){\vector(1,0){1}}
          \put(1.5,7.3){$\gamma^\ast$}
          \put(8.3,7.3){$c$}
          \put(8.3,5.2){$\bar{c}$}
          \put(3.5,-1.5){$(b)$}
        \end{picture}
        }
    \end{picture}
    \setlength{\unitlength}{1pt}
    
  \end{center}
  \caption{\label{fig:Feynman}Two of the four Feynman diagrams
    contributing to the imaginary part of the amplitude for $p +
    \gamma^{\ast} \rightarrow p + c \bar{c}$. The other two are
    obtained by reversing the charge flow of the upper quark line. The
    lower line stands for a constituent quark in the proton as
    explained in \protect\cite{MD:ZP66}, and the dashed lines denote
    nonperturbative gluons.}
\end{figure}
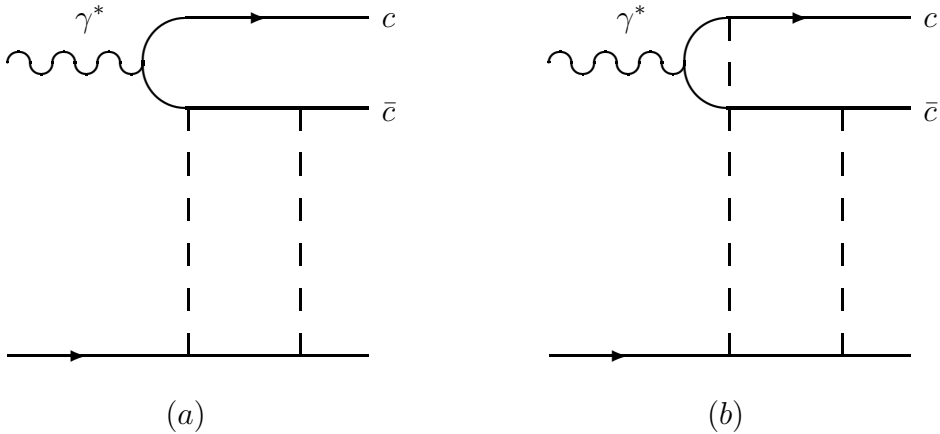

In the high-energy limit the scattering amplitude for our process is
dominated by its imaginary part and we can use the cutting rules to
calculate it. Then the diagrams contributing to \eqref{reaction} are
those in fig.~\ref{fig:Feynman} and the ones obtained by reversing the
charge flow of the upper quark line. In each diagram there is one
off-shell quark, the characteristic scale for its virtuality being
given by \cite{Nikolaev,BarLottWust}
\begin{equation}
\lambda^2 = \frac{\Pt^2 + \mq^2}{1 - \beta}  \eqpt
  \label{BartelsScale}
\end{equation}
For charm production the large quark mass $\mq$ protects this quark
from becoming infrared so that a perturbative treatment of the quark
sector should be safe, even in the photoproduction limit.

The cross section for photons with transverse or longitudinal
polarisation in the $\gp$ frame has been calculated in
\cite{MD:ZP66,MD:angles}. It reads
\newpage
\begin{eqnarray}
  \label{sigma}
  \lefteqn{\frac{d\sigma_{T, L}}{d\Pt^2\, d\Mx^2\, d t} =
    \frac{16}{3} \, \alpha_{\it em} e_c^2 \,
    \frac{\alpha_{s}(\lambda^2)}{\alpha_{s}^{(0)}} \, F_1^2(t) \,
    \xi^{2(1 - \alpha_{\pom}(t))} \cdot } \hspace{5em} \nonumber \\
&& \frac{1}{(\Mx^2 + Q^2)^4} \, \frac{1}{\sqrt{1 - 4 (\Pt^2
    + \mq^{2}) / \Mx^2}} \; {\cal S}_{T, L} \eqcm  
\end{eqnarray}
where $e_c = 2/3$ is the electric charge of the charm quark in units
of the positron charge, and $F_1(t)$ the Dirac form factor of the
proton. We have approximated $t = 0$ for the squared momentum transfer
from the proton, except in $F_1(t)$ and $\alpha_\pom(t)$, which
accounts for most of the $t$-dependence in the cross section
\cite{MD:angles}.  $\alpha_{\pom}(t) \approx 1.085 + t / (2 \GeV)^2$
is the soft pomeron trajectory as observed in hadronic reactions
\cite{beta}. It is introduced by hand in the LN model in order to make
contact with experiment; the approximation of bare two-gluon exchange
(fig.~\ref{fig:Feynman}) would give a factor $\xi^0$ instead of
$\xi^{2(1 - \alpha_{\pom}(t))}$ in the cross section \eqref{sigma}. We
thus assume that the energy dependence of diffractive charm production
is given by the soft pomeron, and furthermore that $\xi$ is the
correct dimensionless variable to be raised to the Regge power $(1
/\xi)^{\alpha_{\pom}(t)}$ in the amplitude. In our numerical
applications we will impose an upper cut of $\xi \le 0.05$ to remain
in a region where the exchange of a pomeron dominates that of
secondary trajectories.

The expressions
\begin{eqnarray}
{\cal S}_{T} &=&
\left(1 - 2\, \frac{\Pt^2 + \mq^{2}}{\Mx^2} \right) \,
  \frac{\Pt^2}{\Pt^2 + \mq^{2}} \, (\Mx^2 + Q^2)^2 \, L_1(\Pt^2,
  \iw)^2 \nonumber \\
 & & + \frac{\mq^{2}}{\Pt^2 + \mq^{2}} \, (\Mx^2 + Q^2)^2 \,
 L_2(\Pt^2, \iw)^2 \nonumber \\  
{\cal S}_{L} &=& 4 \, \frac{Q^{2}}{\Mx^2} \, \frac{\Pt^2 +
  \mq^{2}}{\Mx^2} \,  (\Mx^2 + Q^2)^2 \, L_2(\Pt^2, \iw)^2
  \label{translong}
\end{eqnarray}
in \eqref{sigma} contain integrals $L_1(\Pt^2, \iw)$ and $L_2(\Pt^2,
\iw)$ over the virtuality of the exchanged gluons,
\begin{equation}
L_i(\Pt^2, \iw) = \int_{0}^{\infty} d \lt^2 \,
 [\alpha_{s}^{(0)} D(- \lt^2)]^{2} \, f_{i}(\iv, \iw) \eqcm 
  \hspace{4em}  i = 1,2
  \label{LoopIntegrals}
\end{equation}
with
\begin{eqnarray}
f_1(\iv, \iw) &=& 1 - \frac{1}{2 \iw} \left[ 1 - \frac{\iv + 1 - 2
  \iw}{\sqrt{(\iv + 1 - 2 \iw)^2 + 4 \iw (1 - \iw)}} \right] 
\nonumber \\
f_2(\iv, \iw) &=& 1 - \frac{1}{\sqrt{(\iv + 1 - 2 \iw)^2 + 4 \iw (1 -
  \iw)}}
  \label{functions}
\end{eqnarray}
and\footnote{The definitions of $\iv$ and $\iw$ here differ from
  those in \cite{MD:ZP66}.}
\begin{equation}
  \label{variables}
  \iv = \frac{\lt^2}{\lambda^2} \eqcm \hspace{4em} 
  \iw = \frac{\Pt^2}{\lambda^2} \eqpt
\end{equation}

A simple way to approximate $L_i$ is to Taylor expand the function
$f_i(\iv,\iw)$ about $\iv = 0$ and to keep only the leading term
\begin{equation}
  f_i(\iv, \iw) \approx \iv \cdot \left. \frac{\partial f_i(\iv,
  \iw)}{\partial  \iv} \right|_{\iv = 0}  \eqcm
  \label{LoopApprox}
\end{equation}
after which the integrals reduce to the moment \eqref{moment} of the
gluon propagator. This is however not very good for small $\Pt^2$,
where the variable $\iv$ becomes $\iv = (1 - \beta) \cdot \lt^2 /
m_c^2$ so that one needs a good approximation of the functions
$f_1(\iv)$, $f_2(\iv)$ for $\iv$ from zero to order one. A better
approximation which also leads to the moment \eqref{moment} is
\begin{equation}
  f_i(\iv, \iw) \approx \frac{\iv}{\iv_0} \cdot f_i(\iv_0, \iw)
  \eqcm \hspace{4em} \iv_0 = \frac{l_0^2}{\lambda^2}  \nonumber \\
  \label{ApproxImprove}
\end{equation}
with $\l_0^2 \sim \mu_0^2$. Whereas in \eqref{LoopApprox} one
approximates the curve $f_i(\iv)$ by its tangent at $\iv = 0$ the
approximation \eqref{ApproxImprove} uses instead the line that
intersects the curve at $\iv = 0$ and $\iv = \iv_0$, the corresponding
range in $\lt^2$ from $0$ to $l_0^2$ being the dominant region of
integration in $L_i$. We have varied the parameter $l_0^2$ between
$\mu_0^2 /2$ and $2 \mu_0^2$ and found variations of up to a factor
1.6 in the integrated $\gp$ cross sections $\sigma_T$ and $\sigma_L$
and variations of less than a factor 2.1 for cross sections
differential in $\Mx^2$ or in $\Pt^2$, together with some change in
the shape of the spectra. The effects are stronger at small or zero
$Q^2$ and more pronounced in the $\Pt^2\,$- than in the
$\Mx^2$-spectra. These variations may be seen as reflecting our
uncertainty about the exact shape of the nonperturbative gluon
propagator, which determines the value of $l_0^2$ for which the
approximation \eqref{ApproxImprove} is best. As a benchmark we have
compared the approximated integrals with the exact ones for the model
gluon propagator used in \cite{DL},
\begin{equation}
  D(- l^2) \propto \left[1+\frac{l^2}{(n-1) \mu_{0}^{2}}
  \right]^{-n} \eqcm \hspace{4em} n \ge 4  \eqcm
  \label{SpecialGluon}
\end{equation}
where the proportionality constant can easily be obtained from
\eqref{moment}. With \eqref{ApproxImprove} the errors of the
approximation stay below 10\% for all $\Pt^2$ and $\iw$ we need,
whereas the tangent approximation \eqref{LoopApprox} has errors of
50\% and more when one goes to $\Pt^2 = 0$.

Finally a comment is in order about the value of the strong coupling
in our calculation. In the integral \eqref{moment} it is taken at a
nonperturbative scale given by the dominant virtuality of the
exchanged gluons. However, the first gluon in all diagrams couples at
its upper end to an off-shell quark whose virtuality is in the
perturbative region. We choose to take the coupling for {\em this\/}
vertex at the scale $\lambda^2$ in all four diagrams, and at the small
gluonic scale for the other three vertices.  One might argue that
$\lambda^2$ is the typical scale for the entire upper parts of the
diagrams, and that one should also take the coupling of the second
gluon to the upper quark line at this scale.  Note however that, since
we use the cutting rules, the parts of the diagrams to the left and
the right of the cut lines can be considered independently, and that
the part to the right of the cut is just on-shell quark-quark or
quark-antiquark scattering with no large virtuality involved. We are
aware though that our choice is only a guess.

It is clear that this question of scales leads to an uncertainty in
the normalisation of our predictions. The main problem is not so much
whether $\lambda^2$ is the best choice of a hard scale, which is a
common problem of all leading order calculations. Since the moment
\eqref{moment} includes the nonperturbative coupling $\alpha_s^{(0)}$
we have to multiply the cross section with $\alpha_s(\lambda^2)
/\alpha_s^{(0)}$ if at one of the four quark-gluon vertices we take
the perturbative coupling.  $\alpha_s^{(0)}$ is not well constrained
by phenomenology \cite{Cudell} or theory and other choices than
$\alpha_s^{(0)} = 1$ which we adopt here have indeed been made
\cite{CudellRho}.

\section{Results}
\label{sec:results}

We will now give some numerical predictions for the diffractive
production of charm. We stress once again that we calculate the cross
section for the diffractive final state being a quark-antiquark pair,
which does not include events with $c \bar{c}$ and additional gluons
at parton level. It is also different from $c \bar{c}$-production
through photon-gluon fusion where the gluon is a parton emitted by the
pomeron in the description of Ingelman and Schlein \cite{IS} and where
the final state contains a pomeron remnant. Finally it excludes events
in photoproduction where the photon is resolved.

Let us start with photoproduction. For $W = 220 \GeV$ and $\xi \le
0.05$ we obtain a total rate $\sigma(\gamma p \to c \bar{c} \, p) = 57
\nbarn$. Spectra in $\Pt^2$ and $\Mx^2$ are shown in
fig.~\ref{fig:GammaCharm} and \ref{fig:GammaCharmLog}. The
$\Pt^2\,$-spectrum is well described by a power behaviour
\begin{equation}
\frac{d \sigma_{\gamma p}}{d \Pt^2} \propto (\Pt^2 + m_c^2)^{- \delta}
  \label{photoCharmPt}
\end{equation}
with an exponent $\delta$ between 3.7 and 4.3 in the $\Pt^2\,$-range
of fig.~\ref{fig:GammaCharmLog}. The spectrum of diffractive mass
behaves approximately like $d \sigma_{\gamma p} / d \Mx^2 \propto
\Mx^{- 4.2}$ to the right of its peak.

\newcommand{\spectrum}[4]{
  \begin{picture}(48,55.9)(-0.05,2.5)
    \put(23,57){\large $(#1)$}
    \put(0,50.5){\shortstack[l]{#2}}
    \epsfxsize=0.54\textwidth \put(-5,3){\epsfbox{#3}}
    \put(20,0){#4}
  \end{picture}}

\begin{figure}
  \begin{center}
    \leavevmode
    \setlength{\unitlength}{0.01\textwidth}
    \spectrum{a}{$d \sigma_{\gp} / d \Pt^2$ \\ $[{\rm pb} / {\rm
        GeV}^2$]}{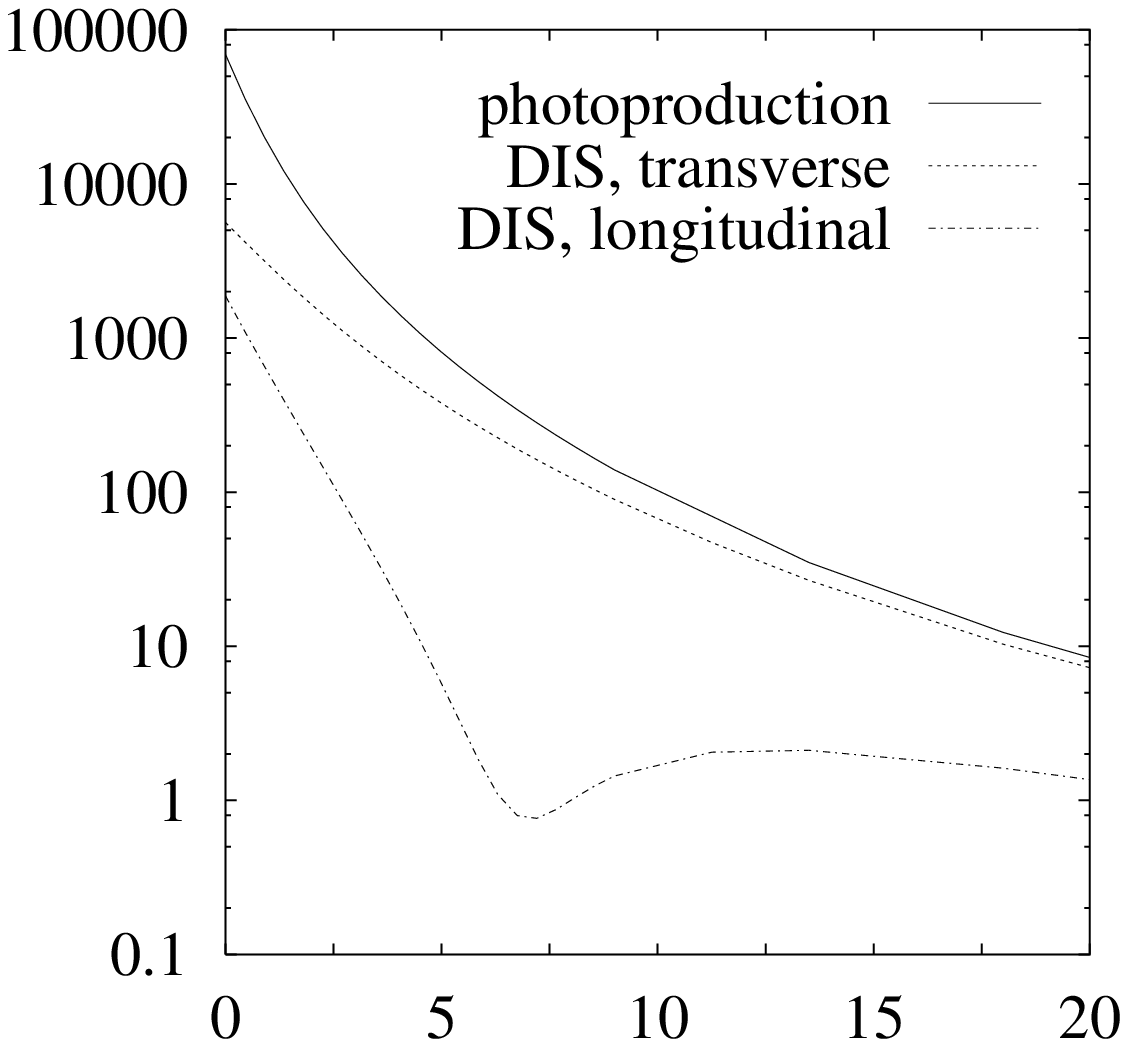}{$\Pt^2 \, [{\rm GeV}^2]$}
    \spectrum{b}{$d \sigma_{\gp} / d \Mx^2$ \\ $[{\rm pb} / {\rm
        GeV}^2$]}{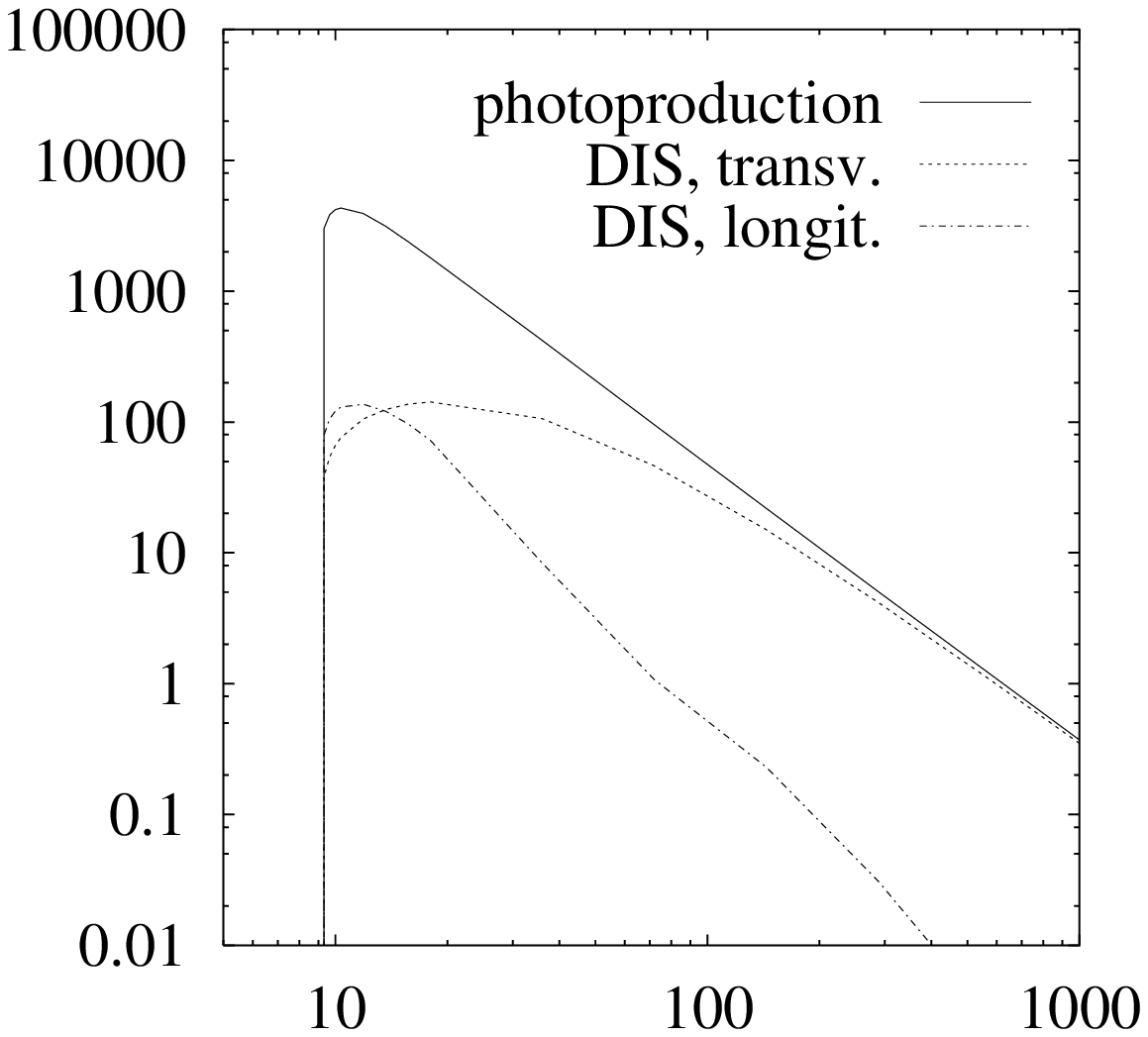}{$\Mx^2 \, [{\rm GeV}^2]$}
    \setlength{\unitlength}{1pt}
  \end{center}
  \caption{\label{fig:GammaCharm}Spectra in $\Pt^2$ and in
    $\Mx^2$ for $\gp \to c \bar{c} \, p$ at $W = 220 \GeV$ with a cut
    $\xi \le 0.05$. The curves are for photoproduction and for
    electroproduction at $Q^2 = 20 \GeV^2$ with transverse or
    longitudinal photons.}
\end{figure}

\begin{figure}
  \begin{center}
    \leavevmode
    \setlength{\unitlength}{0.01\textwidth}
    \begin{picture}(46,52.5)(0,2.5)
      \put(0,48){\shortstack[l]{$d \sigma_{\gp} / d \Pt^2$ \\ $[{\rm
            pb} / {\rm GeV}^2$]}}
      \epsfxsize=0.5\textwidth
      \put(-5,3){\epsfbox{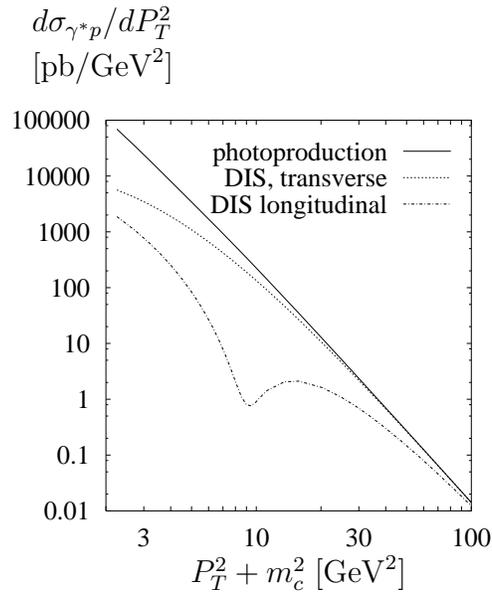}} 
      \put(15,0){$\Pt^2 + \mq^2 \; [{\rm GeV}^2]$}
    \end{picture}
    \setlength{\unitlength}{1pt}
  \end{center}
  \caption{\label{fig:GammaCharmLog}The spectra of
    fig.~\protect\ref{fig:GammaCharm} $(a)$ for a wider range in
    $\Pt^2$, as a function of $\Pt^2 + m_c^2$ in a double logarithmic
    plot. For photoproduction one recognises the power behaviour
    (\protect\ref{photoCharmPt}).}
\end{figure}

Let us move on to electroproduction and first focus on the dependence
on the $\gp$ c.m.\ energy $W$ of the $ep$ cross section, obtained from
the usual relation
\begin{equation}
  d \sigma_{ep} = \frac{\alpha_{\it em}}{\pi} \,
  \frac{d W^2}{W^2} \, \frac{d Q^2}{Q^2} \, \left\{ (1
  - y + y^2 /2) \, d \sigma_T + (1 - y) \, d \sigma_L \right\}
  \label{EpFromGp}
\end{equation}
where we have used the approximation $x \ll 1$. It is determined by
different effects:
\begin{enumerate}
\item the integration element $d W^2 / W^2$ in \eqref{EpFromGp}
\item the $y$-dependent factors multiplying $d \sigma_T$ and $d
  \sigma_L$, note that at $x \ll 1$ one has $y \approx W^2 /s$. They
  decrease with $W$. As they are different for transverse and
  longitudinal photons one cannot calculate their effect without
  knowing the relative contribution of $\sigma_T$ and $\sigma_L$. This
  problem is of course absent in the photoproduction limit.
\item the cut in $\xi$ which ensures that the selected events are
  dominated by pomeron exchange. It leads to an upper limit on the
  diffractive mass $\Mx$ that depends on $W$ and $Q^2$ and whose
  importance is strongest for small $W$ and large $Q^2$. This effect
  can be circumvented by using a fixed cut on $\Mx$ chosen such that
  $\xi$ is always small enough, but at the expense of the total rate
  used in the analysis.
\item the dependence on $W$ of $d \sigma_{T,L} / (d \Pt^2\, d \Mx^2\,
  d t)$. This gives direct information about whether or not this
  process is dominated by the soft pomeron. In our model it comes from
  the factor $\xi^{2(1 - \alpha_{\pom}(t))}$ in \eqref{sigma}, i.e.\ 
  with $W^2 \ll Q^2$ it is $W^{4(\alpha_{\pom}(t) - 1)}$ which is
  quite flat given that the soft pomeron intercept is close to 1.
\end{enumerate}
We find an $ep$ cross section of $120 \pbarn$ for $\sqrt{s} = 296
\GeV$, $\xi \le 0.05$, integrated over $Q^2$ from $7.5 \GeV^2$ to $80
\GeV^2$ and $W$ from $50 \GeV$ to $220 \GeV$.
Table~\ref{tab:BinsCharm} $(a)$ gives the $ep$ cross section in three
different bins of $W$. They are spaced logarithmically to take out the
trivial effect of point 1.\ above. One might also choose the binning
to include the factor $1 - y + y^2 /2$ if one assumes the longitudinal
contribution to the cross section to be small; one then directly
extracts the $W$-dependence of the integrated $\gp$ cross section
$\sigma_T$. The effects of points 2.\ and 3.\ are responsible for the
decrease of the cross section from the second to the third $W$-bin in
our numerical example. Apart from this one sees however clearly the
flat behaviour in $W$ characteristic of soft pomeron exchange.  It
might be useful to analyse experimental data in this way: it focuses
on the $W$-dependence of the $\gp$ cross section and apart from points
2.\ and 3.\ does not involve the details of its dependence on $\Mx^2$
or $Q^2$, so it is relatively model independent. Also it requires only
binning in one variable and thus makes best use of the available
statistics which is likely not to be abundant at HERA. If there were a
strong departure from the soft pomeron energy dependence for these
events it should be seen in such an analysis.

\begin{table}[htbp]
  \caption{\label{tab:BinsCharm}$\sigma_{ep}$ for $ep \to ep\, c
    \bar{c}$ with $\protect\sqrt{s} = 296 \GeV$ and $\xi \le 0.05$.
    $(a)$ For $Q^2$ from $7.5 \GeV^2$ to $80 \GeV^2$ and three
    logarithmically spaced bins in $W$. $(b)$ For $W$ from $50 \GeV$
    to $220 \GeV$ and three logarithmic bins in $Q^2$.}
  \begin{center}
    \leavevmode
    \renewcommand{\arraystretch}{1.2}     
    \begin{tabular}{ccccccc}  
      \multicolumn{7}{c}{ } \\
      \multicolumn{3}{l}{$(a)$} &  &  
      \multicolumn{3}{l}{$(b)$} \\ \cline{1-3} \cline{5-7} 
      \multicolumn{3}{c}{$W [\GeV]$} & & 
      \multicolumn{3}{c}{$Q^2 [\GeV^2]$} \\
      50  to 82  & 82 to 134  & 134  to 220 &
      & 7.5 to 16.5 & 16.5 to 36.3 & 36.3 to 80 \\ \cline{1-3}
      \cline{5-7}  
      39 \pbarn & 44 \pbarn & 40 \pbarn &
      & 73 \pbarn & 36 \pbarn & 14 \pbarn \\ \cline{1-3} \cline{5-7} 
    \end{tabular}
    \renewcommand{\arraystretch}{1}
  \end{center}
\end{table}

In table~\ref{tab:BinsCharm} $(b)$ we give the $ep$ cross section in
logarithmically spaced $Q^2$-bins. Since $y$ is nearly independent of
$Q^2$ at small $x$ this directly shows the $Q^2$-dependence of the
weighted sum of the integrated transverse and longitudinal $\gp$ cross
sections.
 
\begin{figure}
  \begin{center}
    \leavevmode
    \setlength{\unitlength}{0.01\textwidth}
    \spectrum{a}{$d \sigma_{ep} / d \Pt^2$ \\ $[{\rm pb} / {\rm
        GeV}^2$]}{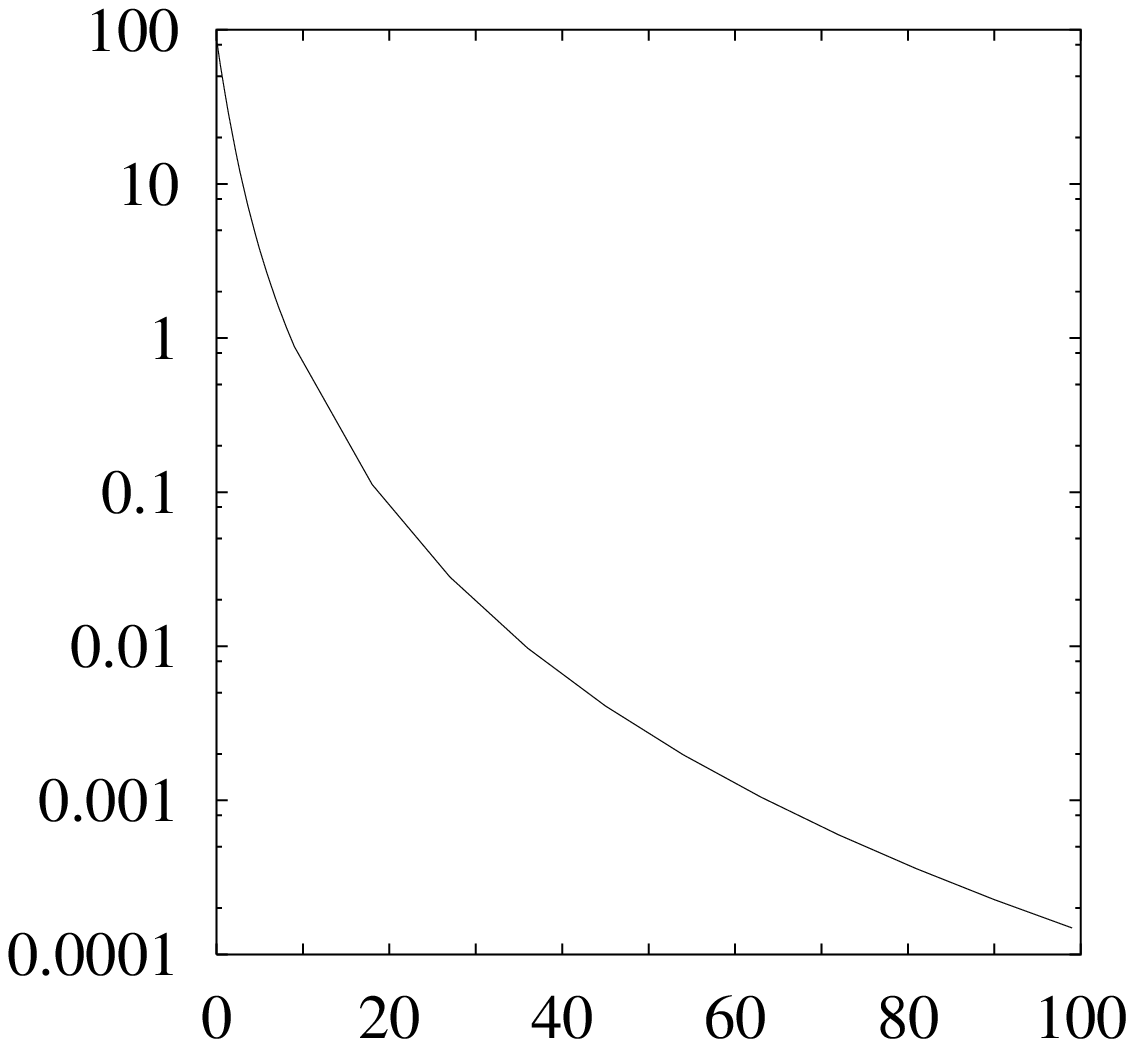}{$\Pt^2 \, [{\rm GeV}^2]$}
    \spectrum{b}{$d \sigma_{ep} / d \Mx^2$ \\ $[{\rm pb} / {\rm
        GeV}^2$]}{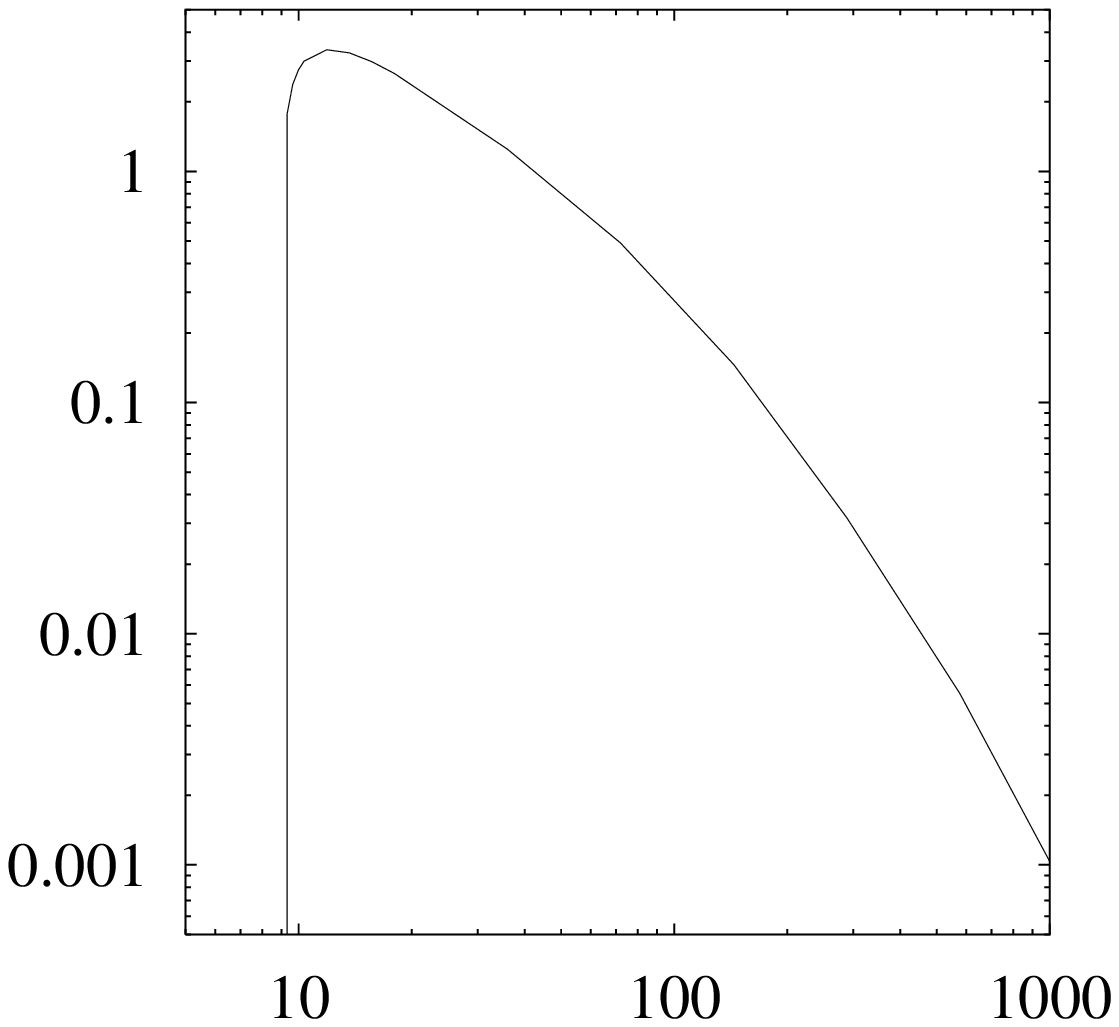}{$\Mx^2 \, [{\rm GeV}^2]$}
    \setlength{\unitlength}{1pt}
  \end{center}
  \caption{\label{fig:EpCharm}Spectra in $\Pt^2$ and in $\Mx^2$
    for $ep \to ep\, c \bar{c}$ at $\protect\sqrt{s} = 296 \GeV$,
    integrated over $\xi \le 0.05$, $W = 50 \GeV$ to $220 \GeV$ and
    $Q^2 = 7.5 \GeV^2$ to $80 \GeV^2$.}
\end{figure}

Examples of $ep$ spectra are shown in fig.~\ref{fig:EpCharm} and $\gp$
spectra for transverse and longitudinal photons in
fig.~\ref{fig:GammaCharm}. In the $\Mx^2$-spectrum of
fig.~\ref{fig:GammaCharm} $(b)$ we can see that the contribution from
longitudinal photons is much smaller than from transverse ones except
at very small $\Mx^2$. This is explained by the factor $(\Pt^2 +
\mq^2) / \Mx^2$ in the differential cross section for longitudinal
polarisation, cf.\ eq.~\eqref{translong}. The $\Pt^2\,$-spectrum for
transverse photons is less steep than in photoproduction at its lower
end, but at large $\Pt^2$ it becomes similar in slope and
normalisation. The dip in the $\Pt^2\,$-spectrum for longitudinal
polarisation is a consequence of a zero in the longitudinal cross
section at fixed $\Pt^2$ and $\Mx^2$ which is due to the behaviour of
the integral $L_2(\Pt^2, \iw)$ in \eqref{LoopIntegrals}. Unfortunately
it becomes completely swamped in the $ep$ spectrum by the transverse
contribution, cf.\ fig.~\ref{fig:EpCharm} $(a)$, so that this effect
is unlikely to be observed.

The $\Mx^2$-spectra can be rewritten in terms of the charm
contribution to the diffractive structure functions $F^{D(4)}_2$,
$F^{D(4)}_T$ and $F^{D(4)}_L$, defined by
\begin{eqnarray}
\frac{d \sigma_{ep}}{d x\, d Q^2\, d \xi\, d t} &=& \frac{4 \pi
  \alpha_{\it em}^2}{x Q^4} 
  \left[ \left( 1 - y + y^2/2 \right) F^{D(4)}_T + 
  (1 - y) \, F^{D(4)}_L \right] \nonumber \\
F^{D(4)}_2 &=& F^{D(4)}_T + F^{D(4)}_L  \eqcm 
  \label{DiffStruct}
\end{eqnarray}
where $T$ and $L$ stand for the contributions of transverse and
longitudinal photon polarisation as usual. These functions depend on
the kinematic variables $x, Q^2, \xi, t$ or, equivalently, on $\xi,
\beta, Q^2, t$. In models with Regge factorisation this dependence
factorises into
\begin{equation}
F^{D(4)}_{T,L}(\xi, \beta, Q^2, t) = f_\pom(\xi, t) \cdot
F^\pom_{T,L}(\beta, Q^2, t) \eqcm
  \label{factorising}
\end{equation}
where in the partonic interpretation of Ingelman and Schlein \cite{IS}
$f_\pom(\xi, t)$ gives the pomeron flux from the proton and
$F^\pom_{T,L}(\beta, Q^2, t)$ are the structure functions of the
pomeron.

Our model has this factorisation property, but this is because it is
put in rather than being one of its predictions: We calculate
two-gluon exchange to leading order in $\xi^{-1}$, so that
$F^{D(4)}_{T,L}(\xi, \beta, Q^2, t)$ depends on $\xi$ via a global
factor, and then we modify the exponent of $\xi$ by hand introducing
the soft pomeron trajectory, which preserves factorisation. Using the
flux factor
\begin{equation}
f_\pom(\xi, t) =  \frac{9 \beta_{0}^{2}}{4 \pi^{2}}
        [F_{1}(t)]^{2}\, \xi^{1 - 2 \alpha_{\pom}(t)}
  \label{DLflux}
\end{equation}
we obtain the curves shown in fig.~\ref{fig:CharmStruct} for the charm
structure functions $F^\pom_T(c\bar{c})$ and $F^\pom_L(c\bar{c})$ of
the pomeron. We observe that the transverse structure function shows
Bjorken scaling for rather small $\beta$. It can be shown from the
expressions \eqref{sigma} and \eqref{translong} that
$F^\pom_T(c\bar{c})$ scales under the condition that the integrated
cross section is dominated by values of $\Pt^2$ that are small
compared with the upper kinematical limit $\Mx^2 /4 - \mq^2$. Because
with our values of $Q^2$ the diffractive mass $\Mx$ is not far from
the production threshold $2 \mq$ for moderate and large $\beta$ this
condition cannot be satisfied in this region and there is no scaling.
Note also that at the value of $\beta$ corresponding to $\Mx = 2 \mq$
both $F^\pom_T(c\bar{c})$ and $F^\pom_L(c\bar{c})$ vanish.

\newcommand{\charmstr}[3]{
  \begin{picture}(48,57)(1,2.5)
    \put(23,54){\large $(#1)$}
    \put(0,51){\shortstack[l]{#2}}
    \epsfxsize=0.54\textwidth \put(-5,3){\epsfbox{#3}}
    \put(19.9,43.2){\small $Q^2 = 8.5 \GeV^2$}
    \put(28,40.5){\small $12 \GeV^2$}
    \put(28,37.8){\small $25 \GeV^2$}
    \put(28,35.1){\small $50 \GeV^2$}
    \put(25,0){$\beta$}
  \end{picture}}

\begin{figure}
  \begin{center}
    \leavevmode
    \setlength{\unitlength}{0.01\textwidth}
    \charmstr{a}{$F^\pom_T(c\bar{c})$}{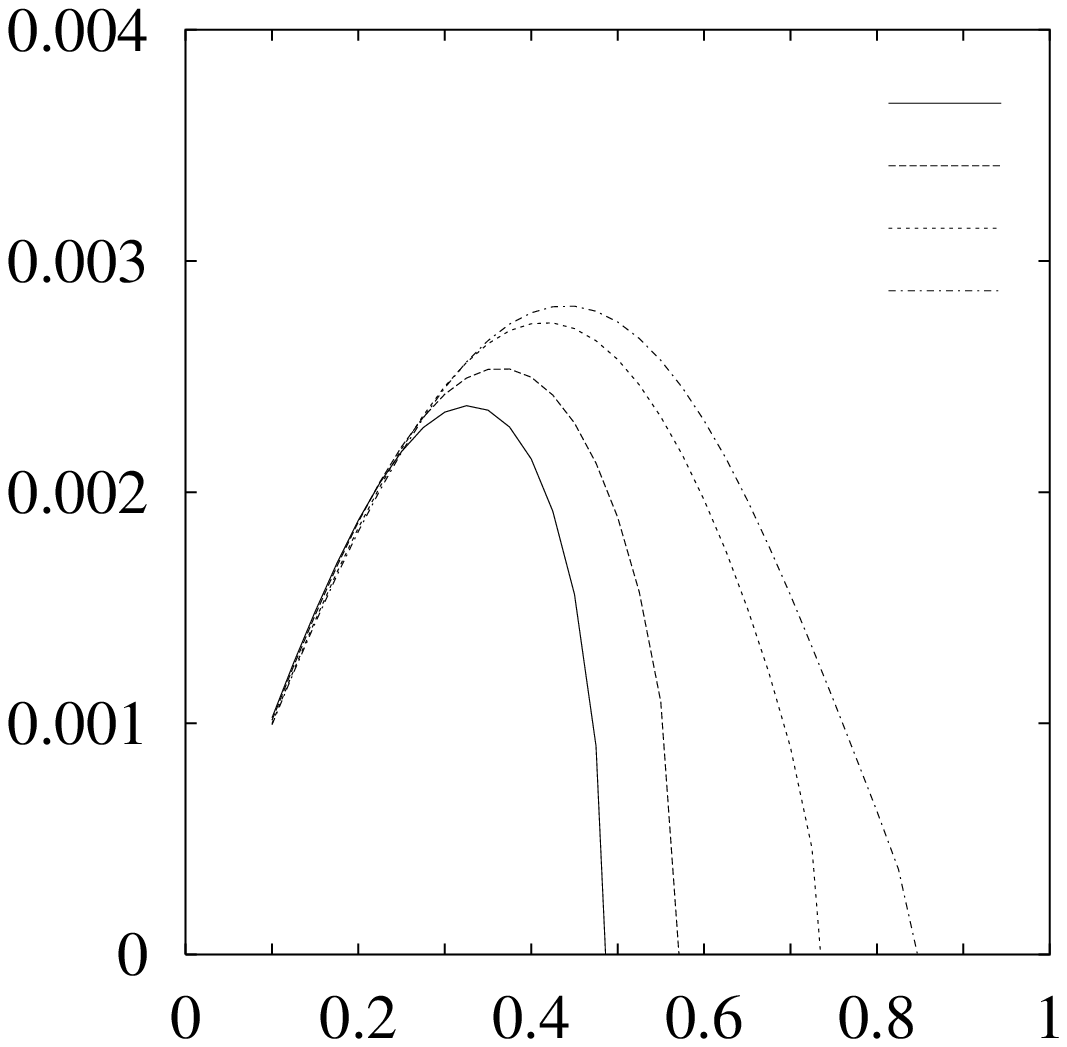}
    \charmstr{b}{$F^\pom_L(c\bar{c})$}{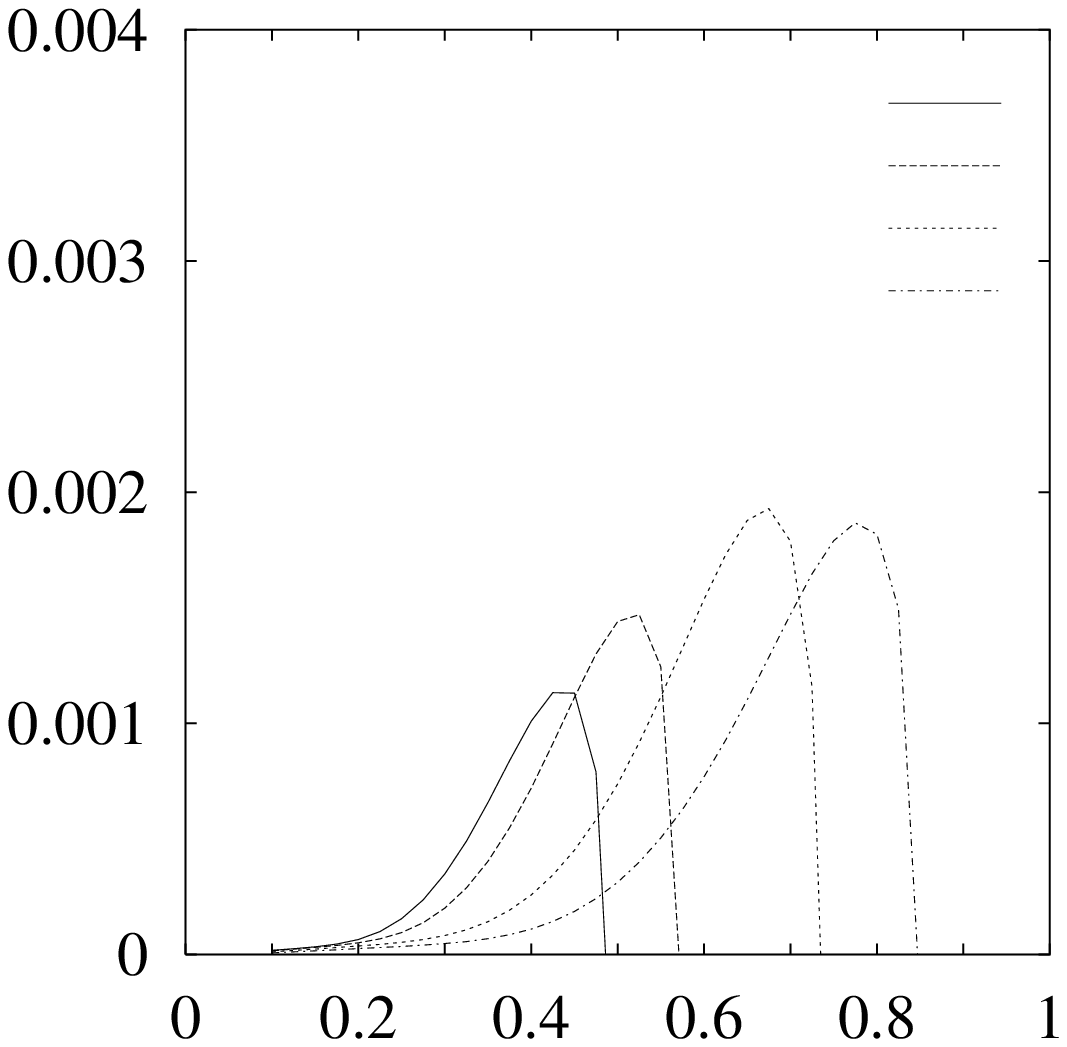}
    \setlength{\unitlength}{1pt}
  \end{center}
  \caption{\label{fig:CharmStruct}Charm structure functions
    $F^\pom_T(c\bar{c})$, $F^\pom_L(c\bar{c})$ of the pomeron for
    transverse and longitudinal photon polarisation at different
    values of $Q^2$.}
\end{figure}

We wish to compare the rate of charm production with the inclusive
diffractive cross section for light flavours $u$, $d$, $s$, which can
also be calculated in our model \cite{MD:ZP66,MD:thesis}. Let us
remark that for light flavours the quark virtuality $\lambda^2$ in
\eqref{BartelsScale} can become small so that one has to assume that a
perturbative treatment of the quarks, possibly with an effective quark
mass of some $100 {\rm \ MeV}$, is good enough for this process. We
find that actually the best description of the HERA data with our
model is obtained when just taking current masses for the $u$, $d$ and
$s$ quarks \cite{MD:thesis}, which we have done in the results to be
shown. For light quarks the approximations \eqref{LoopApprox} or
\eqref{ApproxImprove} of the loop integrals $L_1$, $L_2$ can no longer
be used unless $\Pt^2$ is large so that one has to resort to a
specific form of the gluon propagator $D(l^2)$. The results we show
have been obtained using the model propagator \eqref{SpecialGluon}
with $n = 4$ and freezing the running coupling $\alpha_s(\lambda^2)$
in \eqref{sigma} when it becomes equal to 1. $F^\pom_T$ shows scaling
over the entire $\beta$-range since there is no strong threshold
effect as for charm at large $\beta$. The longitudinal structure
function $F^\pom_L$ is found to behave roughly like $1 / Q^2$ at fixed
$\beta$ and only gives a significant contribution to $F^\pom_2$ when
$\beta$ is large.

Fig.~\ref{fig:HERAcompare} shows the result of our calculation of the
$\xi$-integrated diffractive structure function
$\tilde{F}_2^{D}(\beta, Q^2) = \int_{\xi_{\it min}}^{\xi_{\it max}} d
\xi \int d t\, F^{D(4)}_2(\xi, \beta, Q^2, t)$ together with HERA data
\cite{HERA:struct}. We only show one $Q^2$ per experiment as the
dependence of $F^{D(4)}_2(\xi, \beta, Q^2, t)$ on $Q^2$ is found to be
weak in the data, in good agreement with our results. Remember that
our calculation is at Born level and does not incorporate the effects
of QCD evolution of structure functions.

\newcommand{\heracomp}[4]{
  \begin{picture}(48,67)(1,0)
    \put(23,62.5){\large $(#1)$}
    \put(2,58){\shortstack[l]{#2}}
    \put(27.5,58){\shortstack[l]{#3}}
    \epsfxsize=0.54\textwidth \put(-5,3){\epsfbox{#4}}
    \put(25,0){$\beta$}
  \end{picture}}

\begin{figure}
  \begin{center}
    \leavevmode
    \setlength{\unitlength}{0.01\textwidth}
    \heracomp{a}{$\tilde{F}^D_2$}{$Q^2 = 16
      \GeV^2$}{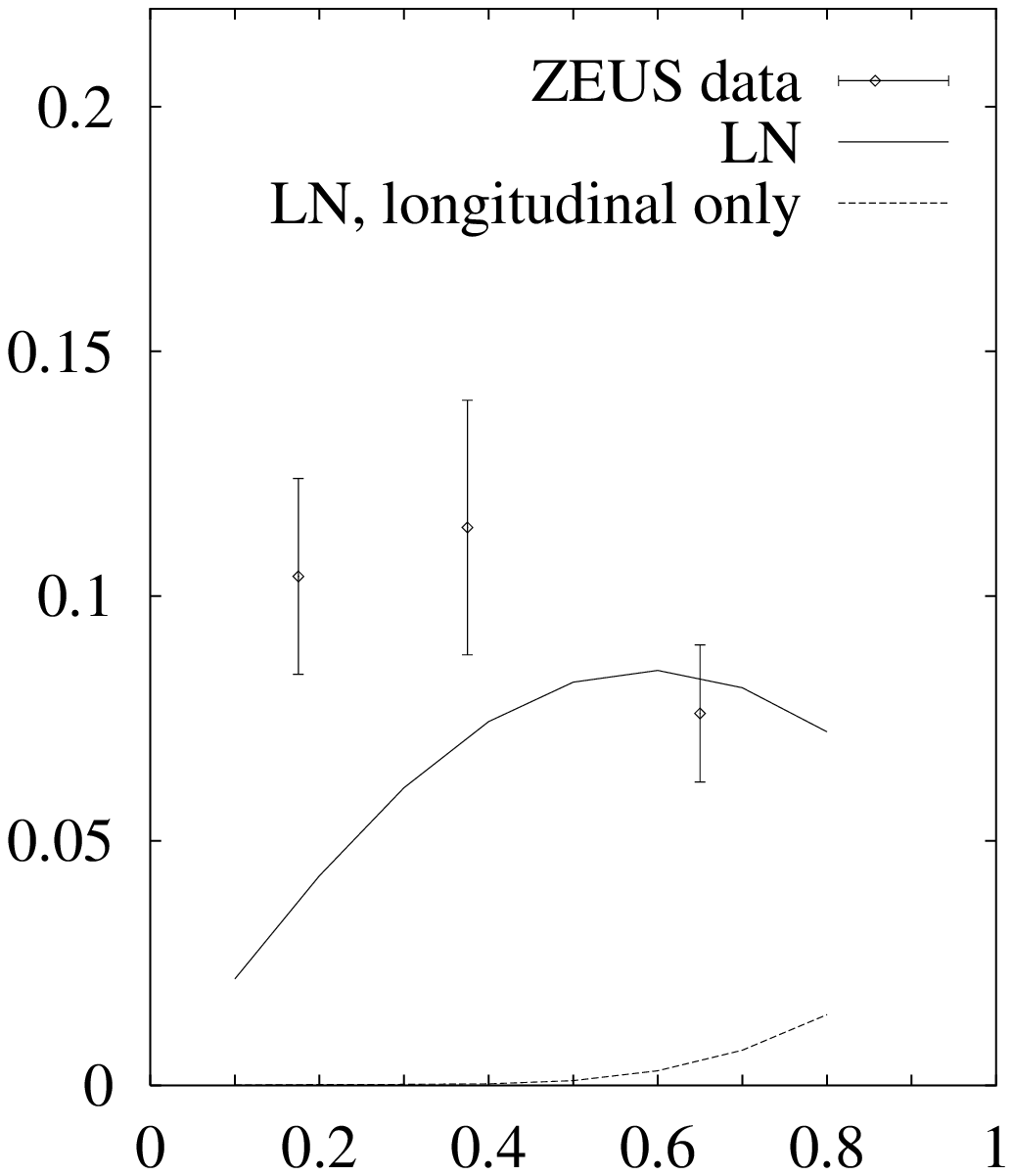} 
    \heracomp{b}{$\tilde{F}^D_2$}{$Q^2 = 25
      \GeV^2$}{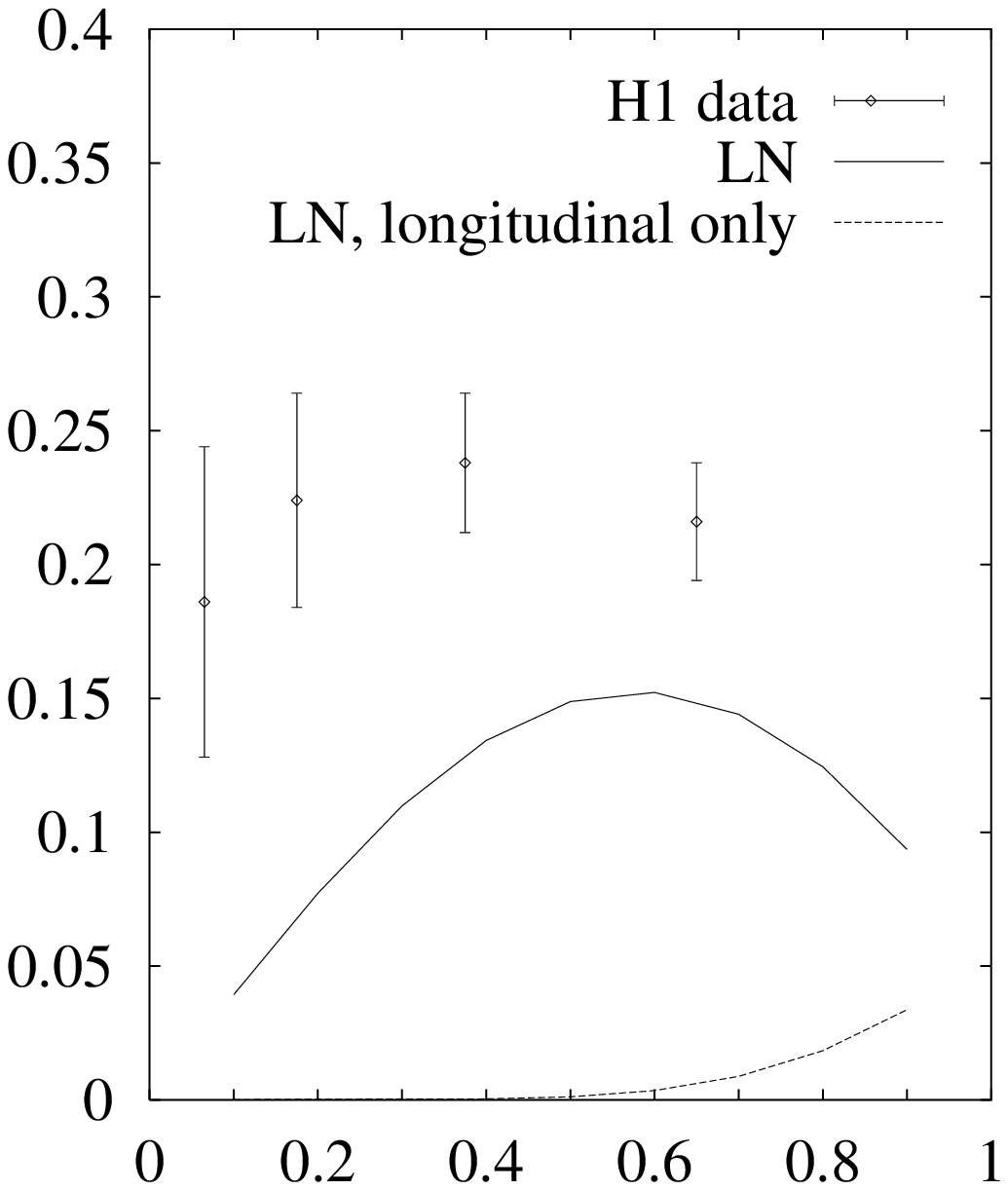} 
    \setlength{\unitlength}{1pt}
  \end{center}
  \caption{\label{fig:HERAcompare}HERA data \protect\cite{HERA:struct}
    for the $\xi$-integrated diffractive structure function
    $\tilde{F}_2^{D}(\beta, Q^2) = \int_{\xi_{\it min}}^{\xi_{\it
        max}} d \xi \int d t\, F^{D(4)}_2(\xi, \beta, Q^2, t)$
    compared with the result in the LN model. Also shown is the
    longitudinal contribution $\tilde{F}_L^{D}$ to $\tilde{F}_2^{D}$.
    Integration limits are $\xi_{\it min} = 6.3 \cdot 10^{-4}$,
    $\xi_{\it max} = 0.01$ in the case of ZEUS and $\xi_{\it min} = 3
    \cdot 10^{-4}$, $\xi_{\it max} = 0.05$ for the H1 data.}
\end{figure}

Regarding the overall normalisation we find that agreement is not too
bad given that we have done a leading order calculation and taking
into account the uncertainty in its normalisation due to the strong
coupling at different scales we discussed in sec.~\ref{sec:calc}. We
also stress that the parameters of our model, $\beta_0^2$, $\mu_0^2$
and $\alpha_s^{(0)}$, have all be determined from pre-HERA data and
that in this sense our prediction is parameter free.  As to the shape
in $\beta$ the data clearly do not show a decrease at small $\beta$ as
our result does. This is not surprising since we only calculate the $q
\bar{q}$-component of the diffractive final state, and at small values
of $\beta$, i.e.\ at large diffractive mass $\Mx$ final states with
additional gluons are expected to be dominant.
Fig.~\ref{fig:HERAcompare} indicates that this might be the case for
values of $\beta$ well above 0.1. We remark that the more recent ZEUS
data \cite{ZEUSstructNew} indicate a rise of $F_2^{(D)4}$ as $\beta$
becomes small whereas the preliminary H1 data \cite{H1structNew} give
a very flat behaviour over the entire $\beta$-range.

In fig.~\ref{fig:CharmStructComp} we compare the predictions of the
model for $F^\pom_2(c \bar{c})$ and $F^\pom_2$ for the three light
flavours, keeping in mind that neither is expected to be a complete
description at small $\beta$. The curves for light quarks are scaled
down by a factor of 20. The fraction of charm comes out quite
small,\footnote{The results for the charm contribution to the
  diffractive structure function presented here are significantly
  smaller than those given in \cite{MD:ZP66} because there we used the
  running coupling $\alpha_s(\Pt^2)$ which unlike
  $\alpha_s(\lambda^2)$ is not limited from above by $\alpha_s(m_c^2)$
  and was frozen when it reached the value 1.} it is not larger than
5\% and decreases with $\beta$.  For $\beta \gsim 0.4$, where we
expect the $q \bar{q}$-component to dominate the final state, we find
the fraction of charm in $F_2^\pom$ to be below 4\%.

\newcommand{\charmstrcomp}[3]{
  \begin{picture}(48,57)(1,2.5)
    \put(23,54.5){\large $(#1)$}
    \put(27.5,50){\shortstack[l]{#2}}
    \epsfxsize=0.54\textwidth \put(-5,3){\epsfbox{#3}}
    \put(28,43.4){\small $F^\pom_2(c \bar{c})$}
    \put(14.5,40.3){\small $0.05 \cdot F^\pom_2(u\bar{u}\, d\bar{d}\,
    s\bar{s})$} 
    \put(25,0){$\beta$}
  \end{picture}}

\begin{figure}
  \begin{center}
    \leavevmode
    \setlength{\unitlength}{0.01\textwidth}
    \charmstrcomp{a}{$Q^2 = 16 \GeV^2$}{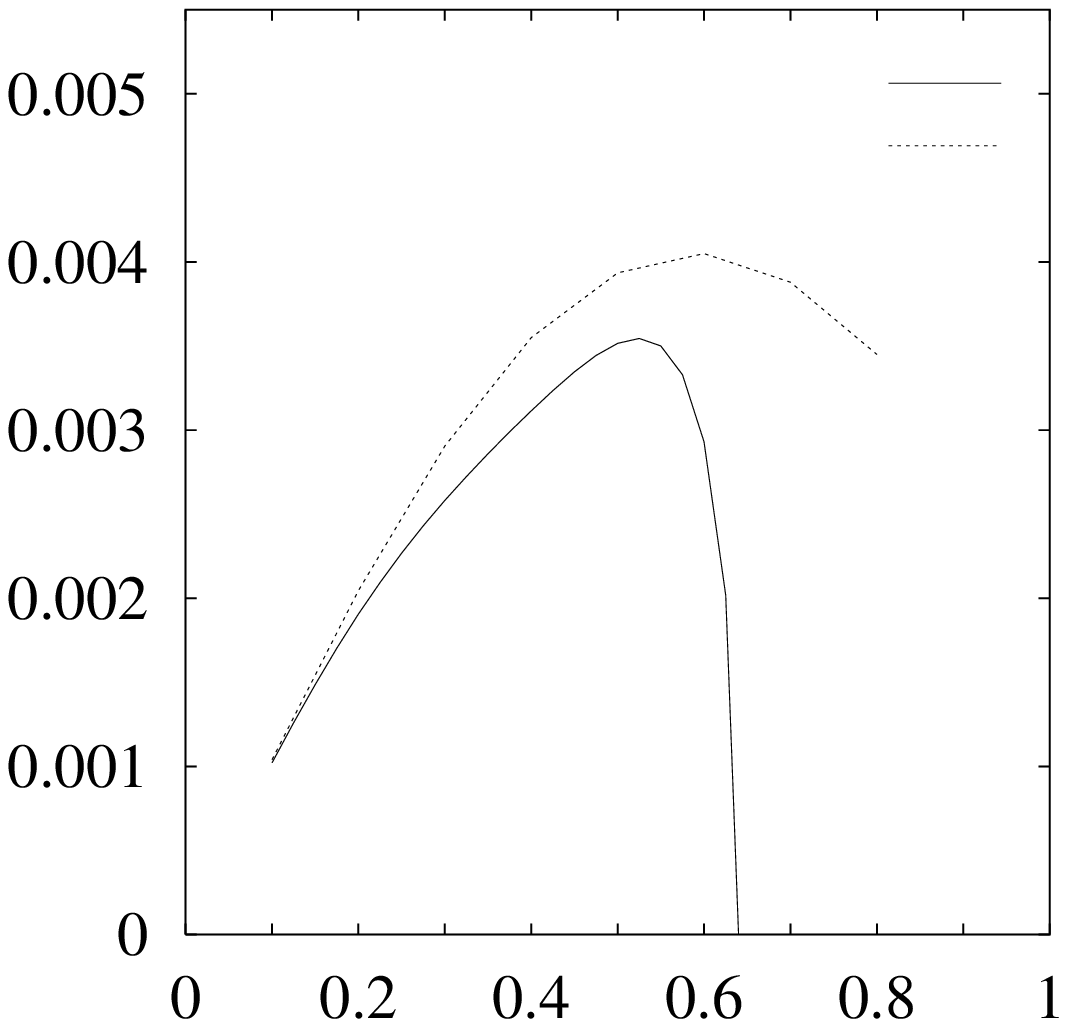}
    \charmstrcomp{b}{$Q^2 = 25 \GeV^2$}{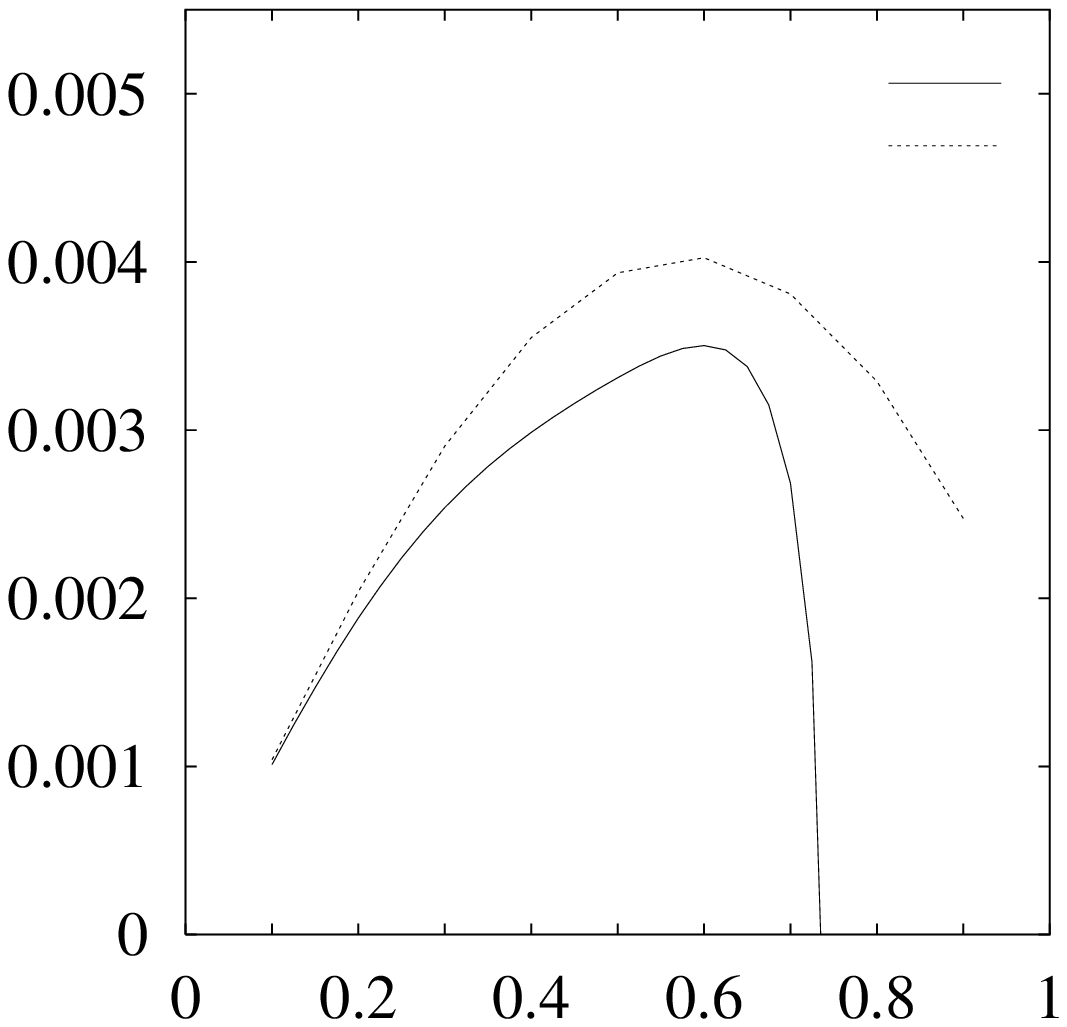}
    \setlength{\unitlength}{1pt}
  \end{center}
  \caption{\label{fig:CharmStructComp}Pomeron structure function
    $F^\pom_2(c\bar{c})$ for charm compared with 0.05 times $F^\pom_2$
    for the three light flavours. $(a)$ for $Q^2 = 16 \GeV^2$ and
    $(b)$ for $Q^2 = 25 \GeV^2$.}
\end{figure}

We observe that for the {\em longitudinal\/} pomeron structure
function the contribution of charm can be considerable. It is largest
for $\beta$ around 0.5 where the ratio between $F^\pom_L(c \bar{c})$
and $F^\pom_L$ for all flavours can reach values close to 0.6. This is
larger than the ratio of the squared electric charges so that apart
from the charge the cross section for charm exceeds that of a light $q
\bar{q}$-pair. Note that the differential longitudinal cross section
has a suppression factor $(\Pt^2 + m_q^2) / \Mx^2$ at small $\Pt^2$
which is less efficient for large quark mass $m_q$. This phenomenon
should however be difficult to observe since the longitudinal part of
$F_2^\pom$ is tiny at $\beta \sim 0.5$ and only visible at rather high
$\beta$ and not too large $Q^2$, where $\Mx^2$ is below the charm
threshold.

To conclude we wish to report an observation regarding the ratio of
the rates for diffractive $c \bar{c}\,$- and $b \bar{b}\,$-production.
We compare the integrated cross sections for photoproduction, so that
the scale $Q^2$ cannot influence this ratio, taking $W = 220 \GeV$ and
imposing $\xi \le 0.05$. We checked that at this $W$ the effect of the
phase space reduction through the cut in $\xi$ is about the same for
charm and bottom: calculating the cross sections for very high $W$
where the $\xi$-cut has no effect we verified that the increase of the
cross sections is almost entirely due to the factor $W^{4(\alpha_\pom
  -1)}$ from pomeron exchange. With $m_c = 1.5 \GeV$ and $m_b = 4.5
\GeV$ we find that the cross section for bottom production is about
440 times smaller than for charm, implying that this process should be
quite impossible to observe at HERA. Taking out the squared electric
charges and the running strong coupling, which for simplicity we take
here at fixed scales $m_c^2$ or $m_b^2$, we obtain
\begin{equation}
\frac{\sigma(\gamma p \to c \bar{c}\, p)}{\sigma(\gamma p \to b
    \bar{b}\, p)} \cdot \frac{e_b^2 \,\alpha_s(m_b^2)}{e_c^2
    \,\alpha_s(m_c^2)} \approx 71 \approx \left( \frac{m_b}{m_c}
    \right)^{3.9}  \eqpt 
  \label{BottomCharm}
\end{equation}
Apart from the effect of the running coupling the integrated cross
section appears to scale with the quark mass approximately like $1 /
m_q^4$. To check this we have calculated the photoproduction cross
section as a function of $m_q$ between $m_q = 1.5 \GeV$ and $15 \GeV$,
for very large $W$ so that the $\xi$-cut has no effect. Dividing by
the running coupling $\alpha_s(m_q^2)$ we find indeed an approximate
power behaviour in $1 / m_q$ with an exponent between 3.6 and 4.2. A
behaviour in $1 / m_q^4$ looks like the effect of the off-shell
propagators in the amplitude whose denominators are limited by
$m_q^2$. Notice that the numerators of the propagators, which also
contain one power of $m_q$ do not seem to enter in the same way, their
role is more complicated because the numerator of a quark propagator
has a Dirac matrix structure.

\section{Summary}
\label{sec:sum}

We have calculated diffractive production of a $c \bar{c}$-pair in
$\gp$ collisions with real or virtual photons in the model of
nonperturbative two-gluon exchange due to Landshoff and Nachtmann.
This allowed us to give numerical predictions for diffractive charm
production at HERA, for cross sections and spectra in $\Mx^2$ and
$\Pt^2$. 

In photoproduction we find a $\gamma p$ cross section in the region of
$60 \nbarn$ for $W$ around $200 \GeV$. The mass spectrum peaks at
rather low values of $\Mx$ and then falls off roughly like $1 /
\Mx^4$, whereas the spectrum of the transverse momentum approximately
behaves like a power of $\Pt^2 + \mq^2$ with an exponent around $-4$.

In diffractive DIS the $ep$ cross section we obtain is of order $100
\pbarn$ for $Q^2$ from $7.5 \GeV^2$ to $80 \GeV^2$ and $W$ from $50
\GeV$ to $220 \GeV$. The $\gp$ cross sections exhibit a flat
dependence on $W$, typical of models with soft pomeron exchange. We
suggest that even a coarse logarithmic binning in $W$ of the $ep$
cross section should be useful to test this in the data.

Expressing the $c \bar{c}$ mass spectra in terms of the diffractive
charm structure functions we find that the transverse contribution
$F^{D(4)}_T(c\bar{c})$ does not scale for $\beta \gsim 0.3$ due to the
restricted phase space at HERA values of $Q^2$. The longitudinal
structure function $F^{D(4)}_L(c\bar{c})$ for charm is comparable to
the transverse one at small diffractive mass $\Mx$, i.e.\ close to the
largest kinematically allowed $\beta$ at given $Q^2$, at lower $\beta$
it is negligible.

We have then compared $F^{D(4)}_2(c\bar{c})$ with the diffractive
structure function $F^{D(4)}_2$ for light flavours, calculated in the
same model, which reproduces the HERA data within a factor of 2 or so,
except in the region of small $\beta$ where the neglect of final
states other that $q \bar{q}$ becomes a bad approximation. The
contribution of $c \bar{c}$ to the diffractive structure function
$F^{D(4)}_2$ comes out below 5\% in our model.

The strong quark mass dependence of the integrated $q \bar{q}$ cross
section can be understood in a simple way from the denominator of the
propagator for the off-shell quark in the Feynman diagrams, which
appears squared in the cross section. Its typical value is given by
$\langle \lambda^2 \rangle = (\langle \Pt^2 \rangle + m_q^2) / (1 -
\beta)$, where $\langle \Pt^2 \rangle$ is some average $\Pt^2$. For
photoproduction of heavy quarks we find indeed a quark mass dependence
of the cross section approximately like $1 / m_q^4$, which together
with the effect of the running strong coupling and the different quark
charges leads to a ratio of $b \bar{b}\,$- to $c \bar{c}\,$-production
of $1 /440$.

\subsection*{Acknowledgements}

I am grateful to P V Landshoff for his continued interest in this work
and for many conversations. Thanks are also due to I T Drummond and H
Lotter for discussions. This work was initiated by the DESY workshop
on ``Future Physics at HERA'' and I would like to thank the conveners
of the working group on ``Hard diffractive processes'', H Abramowicz,
L Frankfurt and H Jung, for the stimulating and pleasant atmosphere
during the workshop.

I acknowledge the financial support of the EU Programme ``Human
Capital and Mobility'', Network ``Physics at High Energy Colliders'',
Contracts CHRX-CT93-0357 (DG 12 COMA) and ERBCHBI-CT94-1342.


\begin{thebibliography}{99}

\newcommand{\NP}[1]{Nucl.~Phys.\ B#1}
\newcommand{\PL}[1]{Phys.~Lett.\ B#1}
\newcommand{\PR}[1]{Phys.~Rev.\ D#1}
\newcommand{\PRL}[1]{Phys.~Rev.~Lett.\ #1}
\newcommand{\ZP}[1]{Z.~Phys.\ C#1}

\bibitem{HERA:discover} ZEUS Collaboration, \PL{315} (1993) 481;\\ H1
  Collaboration, \NP{429} (1994) 477
  
\bibitem{HERA:workshop} G Briskin and M F McDermott, ``Diffractive
  Structure Functions in DIS'' in: \textit{Future Physics at HERA},
  Proc.~of the workshop 1995/96, eds.~G Ingelman, A De Roeck and
  R Klanner, DESY 1996
  
\bibitem{GehrStir} T Gehrmann and W J Stirling, \ZP{70} (1996) 89
  
\bibitem{Nikolaev} M Genovese, N N Nikolaev and B G Zakharov,
  \PL{378} (1996) 347

\bibitem{Teubner} E M Levin, A D Martin, M G Ryskin and T Teubner,
  Durham preprint DTP-96-50, hep-ph/9606443

\bibitem{Lotter} H Lotter, DESY--96--260, hep-ph/9612415

\bibitem{LN} P V Landshoff and O Nachtmann, \ZP{35} (1987) 405

\bibitem{DL} A Donnachie and P V Landshoff, \NP{311} (1988/89) 509
  
\bibitem{Cudell} J R Cudell, A Donnachie and P V Landshoff, \NP{322}
  (1989) 55

\bibitem{BarLottWust} J Bartels, H Lotter and M W\"usthoff, \PL{379}
  (1996) 239

\bibitem{MD:ZP66} M Diehl, \ZP{66} (1995) 181
  
\bibitem{MD:angles} M Diehl, Palaiseau preprint CPTH-S472-1096,
  hep-ph/9610430

\bibitem{beta} A Donnachie and P V Landshoff, \NP{267} (1986) 690

\bibitem{CudellRho} J R Cudell, \NP{336} (1990) 1

\bibitem{IS} G Ingelman and P Schlein, \PL{152} (1985) 256
  
\bibitem{MD:thesis} M Diehl, ``Diffraction in electron-positron
  collisions'', Ph.~D. thesis, University of Cambridge, 1996
  (unpublished)

\bibitem{HERA:struct} H1 Collaboration, \PL{348} (1995) 681;\\ ZEUS
  Collaboration, \ZP{68} (1995) 569

\bibitem{ZEUSstructNew} ZEUS Collaboration, \ZP{70} (1996) 391

\bibitem{H1structNew} H1 Collaboration, S Tapprogge et al.,
  ``Diffractive deep inelastic scattering'', Talk given at the 31st
  Rencontres de Moriond: QCD and high-energy hadronic interactions,
  Les Arcs, France, 23--30 March 1996, hep-ex/9605007

\end{thebibliography}
\end{document}